\documentclass[12pt]{article}

\usepackage[totalheight = 23cm, totalwidth = 17cm]{geometry}
\usepackage{amssymb,amsmath,amsfonts,amsbsy,graphicx}
\usepackage{color}

\def\mathbi#1{\textbf{\em #1}}

\newcommand{\calH}{{\cal H}}
\newcommand{\calL}{{\cal L}}
\newcommand{\calO}{{\cal O}}
\newcommand{\calR}{{\cal R}}

\begin{document}

\begin{titlepage}

\rightline{\footnotesize{APCTP-Pre2015-010}}

\begin{center}

\vskip 1cm

\LARGE{\bf Breaking discrete symmetries \\ in the effective field theory of inflation}

\vskip 1cm

\large{
Dario Cannone$^{a,b}$, \,
Jinn-Ouk Gong$^{c,d}$ \, and \,
Gianmassimo Tasinato$^{e}$
}

\vskip 0.5cm

\small{\it
$^a$Dipartimento di Fisica e Astronomia ``G. Galilei'', Universit\`a degli Studi di Padova,
\\
I-35131 Padova, Italy
\\
$^b$INFN, Sezione di Padova, I-35131 Padova, Italy
\\
$^c$Asia Pacific Center for Theoretical Physics, Pohang 790-784, Korea 
\\
$^d$Department of Physics, Postech, Pohang 790-784, Korea
\\
$^e$Department of Physics, Swansea University, Swansea SA2 8PP, UK
}

\vskip 1.2cm

\end{center}

\begin{abstract}

We study the phenomenon of discrete symmetry breaking during the  inflationary 
epoch, using a model-independent approach based on the effective field theory 
of inflation. We work in a context where both time reparameterization symmetry
and spatial diffeomorphism invariance can be broken during inflation. 
We determine the leading derivative operators in the quadratic action for 
fluctuations that break parity and time-reversal. 
Within suitable approximations, we study their consequences for the dynamics 
of linearized fluctuations. 
 Both in the scalar  and tensor sectors, we show that such operators can lead to  new 
direction-dependent phases for  the modes involved. 
  They do not affect the power spectra, 
but can  have consequences for higher correlation 
functions. Moreover, a small quadrupole contribution to the  sound speed
can be generated.

\end{abstract}

\end{titlepage}

\setcounter{page}{0}
\newpage
\setcounter{page}{1}

 \section{Introduction}
 \label{sec:intro}

The inflationary paradigm~\cite{inflation} provides a convincing explanation for the   
statistical properties of the temperature fluctuations in the cosmic microwave 
background (CMB) and of the distribution of galaxies on large scale  in our universe.
The generic  predictions of the simplest models of inflation, based on a single scalar 
field slowly rolling down on a potential, fit very well current 
observations~\cite{Ade:2015lrj}. On the other hand, the wealth of available 
observational data motivates a detailed theoretical analysis of inflationary scenarios. 
Indeed, more refined theoretical predictions -- when tested by observations --  allow 
us to characterize more accurately the mechanism of inflation.

The effective field theory (EFT) of inflation~\cite{Cheung:2007st,Weinberg:2008hq}
has emerged in recent years as a powerful unifying approach for analyzing broad 
classes of inflationary scenarios; see~\cite{Tsujikawa:2014mba} for a review.
After specifying general properties of the system, such as the number of available 
degrees of freedom and the symmetries respected (or slightly broken) by the fields 
that drive inflation, one can write a general, model-independent action governing  
cosmological fluctuations\footnote{There is another, top-down EFT which results
from integrating out heavy degrees of freedom given a mother theory~\cite{topEFT}.}. 
This action is characterized by operators that can be associated with the amplitude 
and scale dependence of the correlation functions of the curvature and tensor 
perturbations. Measuring such cosmological observables can allow us to test broad 
classes of scenarios associated with the operators that constitute the effective action 
for inflationary fluctuations. Moreover, a systematic analysis of all possible operators 
compatible with the given symmetries can also suggest new effects, not predicted or 
analyzed within the specific models studied so far, that if supported by observations 
would motivate new directions for model building.

During inflation, time translation invariance is broken by the homogeneous 
cosmological background. Motivated by this, most of the works on the EFT of inflation 
until now  has  focussed on analyzing setups where the dynamics of perturbations breaks 
time diffeomorphisms, while maintaining spatial diffeomorphism invariance.
On the other hand, we know very little about the nature of the fields driving the 
inflationary epoch. The simplest case  is a single scalar field breaking time
reparameterization only. But there is  also the possibility of inflationary systems 
that break also spatial diffeomorphism invariance, like inflationary vector 
fields~\cite{Golovnev:2008cf} (see e.g.~\cite{Maleknejad:2012fw,Soda:2012zm} for 
review), or alternatively a set of scalar fields obeying special internal symmetries
 that acquire space-dependent vacuum expectation values,  
as in the case of solid/elastic inflation~\cite{Endlich:2012pz,Gruzinov:2004ty} 
(see also~\cite{solid/elastic}). Motivated by these considerations, 
we find interesting to exploit the EFT of inflation to explore more general options  
than those studied so far, also considering the effects of slightly broken spatial 
diffeomorphism invariance. Such effective description, as we will see, also allows one 
to describe systems where spatial anisotropies can be generated during inflation, 
avoiding Wald's no-hair theorem~\cite{Wald:1983ky}. The power of EFT is that 
we do not need to commit on specific models to develop our arguments, that are 
based on general properties of the action of perturbations, and that allow us to explore 
generic cosmological consequences of systems that break all diffeomorphisms 
during inflation. The aim is to identify in a model-independent way novel operators 
that can lead to new effects associated with inflationary observables, as  
non-standard correlations among inflationary perturbations.

The analysis of such systems was  started in~\cite{Cannone:2014uqa} -- building on 
the results of~\cite{Endlich:2012pz,Blas:2009my} -- where it has been shown that, when spatial 
diffeomorphism invariance is broken, the EFT of inflation allows for operators that 
modify the usual dispersion relations in the scalar and tensor sectors. They can 
lead to blue spectrum in the tensor sector, and can induce non-vanishing 
anisotropic stress in the scalar sector that violates the conservation of the curvature 
perturbation on large scales. See also~\cite{otherEFT}.

In this article we instead focus on the interesting class of operators  that break 
{\it discrete symmetries} during inflation. 
  In particular, using the EFT of inflation, 
we consider the consequences of the leading operators that break discrete 
symmetries as parity and time-reversal during this epoch. Unless discrete 
symmetries are imposed by hand on the theory under consideration, such 
operators will normally be generated by inflationary dynamics or renormalization  effects: 
hence it is interesting  to explore their consequences. Parity-violating interactions
have been studied in great detail for their consequences in the CMB, starting 
with~\cite{Lue:1998mq}. These operators can be   associated with the amplification 
of the amplitude of one of the circular polarization of tensor modes around horizon 
crossing, leading to distinctive effects associated with $TB$ and $EB$ 
cross correlations in the CMB~\cite{TB-EB}. Moreover, parity-violating operators 
can also affect the scalar sector, leading to statistical anisotropies 
in the bispectrum, or also explain anomalies in the CMB. 
The realization of models leading to parity violation during inflation and their   
observational consequences have been motivated by, for example, 
pseudoscalars coupled to gauge fields~\cite{Carroll:1998zi}
or Chern-Simons modifications of  gravity~\cite{CSgravity}. 
See e.g. \cite{bigref} for a selection of papers discussing both theoretical and 
observational aspects of parity violation during inflation.

To the best of our knowledge, there are no studies on the consequences of 
operators that contain a single derivative of time coordinate: in absence of better
name, we say that these operators  break time-reversal symmetry during inflation. 
In this article, using the model-independent language of the EFT of inflation, 
we study selected operators that break the aforementioned discrete symmetries,  
and their effects for the dynamics of linearized perturbations around a homogeneous 
and isotropic Friedmann-Robertson-Walker (FRW) universe.  
Both in the scalar and tensor sectors, we show that such operators can lead to  new 
direction-dependent phases for the  fluctuations, and moreover
 a small quadrupole contribution to the effective sound speed.  A direction-dependent phase 
does not affect the power spectrum, but can have consequences for higher 
correlation functions.

We start in Section~\ref{sec-sys} describing the system that we consider, and 
on how to describe breaking of spatial diffeomorphisms during inflation using 
an EFT approach. When spatial diffeomorphisms are broken, background 
spatial anisotropies can be present during inflation, both in the metric and the 
energy-momentum tensor (EMT). Then we establish a perturbative scheme 
based on the hypothesis that such anisotropies are small, and find general 
conditions to eliminate anisotropies from the background metric while 
maintaining anisotropies in the EMT. In Section~\ref{sec-qua} we discuss the 
leading EFT operators, quadratic in perturbations, that break discrete symmetries
during inflation, and their general features. In  Section~\ref{sec-lin} we discuss 
their consequences for the dynamics of  fluctuations at the level of linearized 
equations of motion. In Section~\ref{sec-con} 
we present our conclusions, followed by three technical appendices.

\section{System under consideration}
\label{sec-sys}

We use an EFT approach to  inflation in a setup that does not necessarily 
preserve spatial diffeomorphism invariance. Such a scenario can be realized  
if the fields that drive inflation acquire vacuum expectation values ({\it vevs}) 
characterized by spatially non-trivial profiles. This situation can still be 
compatible with a good degree of spatial homogeneity and isotropy at the 
background level, if the system has  additional internal symmetries, or if the 
aforementioned {\it vevs} are small or combine in such a special way to 
mostly cancel the background anisotropies. Explicit examples of such 
situations are solid inflation -- where scalars acquire space-dependent
 {\it vevs}  or models in which vector fields drive inflation -- that can select preferred directions.
On the other hand, the EFT approach that we adopt allows us not 
to commit on a specific model for developing our arguments.

We start discussing a generalized version of the unitary gauge condition 
during inflation, that is useful for characterizing the dynamics of fluctuations.  
In such a gauge, all fluctuations are limited to those in the metric. 
We then continue discussing the background configuration that we adopt,  
using the language of EFT of inflation. We establish a perturbative scheme 
that allows us to solve the equations of motion for a general anisotropic 
background solution, in a setup that potentially  breaks all diffeomorphisms and that can 
violate the hypothesis of Wald's no-hair theorem. Within our 
approximations, we then determine conditions that allow  us  to establish spatial 
isotropy and homogeneity in the metric. On the other hand  the EMT can contain fields 
that select a preferred spatial direction, and this will be important for characterizing
operators that break discrete symmetries.

\subsection{Unitary gauge condition with broken space-time diffeomorphisms}
\label{sec-sf}

We start discussing a generalized version of single field inflation condition, in a 
context where  time and spatial diffeomorphisms can be simultaneously broken. 
A diffeomorphism transformation acts on the coordinates of space-time as
\begin{equation}\label{fulldi}
x^\mu \to x^\mu + \xi^\mu(t,\mathbi{x})
\end{equation}
for an arbitrary function $\xi^\mu = (\xi^0,\xi^i)$. In conformally flat space-time, 
the function $\xi^i$ can be decomposed into a transverse vector and a longitudinal 
scalar as $\xi^i = \xi_T^i + \partial^i \xi_L$ with $\partial_i \xi^i_T = 0$.

Inflation corresponds to a period of quasi-de Sitter expansion of the universe. 
Usually, the source driving inflation breaks the time reparameterization invariance. 
For example, in models where  a single scalar field 
$\phi(t,\mathbi{x}) = \phi_0(t) + \delta\phi(t,\mathbi{x})$
drives inflation, the time diffeomorphism acts non-linearly on the inflaton perturbation: 
\begin{equation}
\label{trphi0}
\delta \phi \to \delta\phi - \dot{\phi}_0\xi^0 \, .
\end{equation}
Since the scalar perturbation transforms non-trivially under diffeomorphisms, it is not
gauge invariant. A gauge invariant quantity can be formed by combining it with metric
perturbations that transform under time reparameterization. A gauge can then be selected
 -- the so-called {\em unitary gauge} -- where $\delta \phi = 0$ 
and all dynamical degrees of freedom are stored in the metric perturbations.  
In this gauge, the effective Lagrangian 
 for fluctuations contains metric perturbations only, that break explicitly time reparameterization, and
 leave spatial diffeomorphisms unbroken.
  See~\cite{Cheung:2007st} for 
details. Such a scenario is generally called ``single-field'' inflation since there is 
a unique field driving the inflationary expansion, and its background value can be 
unambiguously used as inflationary clock.

However, we do not really know what was occurring during inflation, and it could very 
well be that the system of fields driving inflation was richer than a single scalar field, 
as explained in Section~\ref{sec:intro}. To be more general, we consider the case 
for inflation driven by a set of fields, which we collect within a four-vector $\Psi^\mu$. 
 We can consider a case where the background values $\bar{\Psi}^{\mu}$ depends not only  on time,    
 but they are allowed to depend also on the spatial coordinates. 
The perturbation $\delta\Psi^\mu$ transforms under the full diffeomorphism 
transformation \eqref{fulldi} as
\begin{equation}
\delta\Psi^\mu \to \delta\Psi^\mu - \bar{\Psi}^{\mu}_{,\,\nu} \xi^\nu \, .
\end{equation}
We assume that the background value  $\bar{\Psi}^{\mu}$
has a suitable profile so  that -- 
 by choosing appropriately the gauge function $\xi^\mu$ -- 
a  gauge can be selected, where all field perturbations can be consistently set to zero: 
$\delta \Psi^\mu = 0$. In this gauge  the propagating degrees of freedom are limited to the
metric fluctuations only. This will constitute the  {\it generalized unitary gauge field 
condition} for the setup of fields we consider in this article. A special case of 
such condition, for broken spatial diffeomorphisms only  was discussed 
in~\cite{Endlich:2012pz}. Our gauge condition generically breaks {\it time} 
as well as {\it spatial} diffeomorphism invariance: the resulting action for metric 
fluctuations consequently breaks all diffeomorphisms\footnote{This generalized 
single field condition, among other things, implies that interaction among the 
Goldstone bosons of the broken diffeomorphism invariance can be constrained 
by non-linearly realized symmetries. The latter will lead to consistency conditions 
for correlation functions that generalize the ones found in the ``standard'' case 
where only time diffeomorphisms are broken. This interesting subject will be 
developed elsewhere.}. 
Notice that our generalized single field condition does {\it not} imply that 
inflation is driven by a single background field, and perturbations are not 
necessarily adiabatic, and anisotropic stress
can arise.  This has been discussed in the explicit realization              
in~\cite{Endlich:2012pz,Cannone:2014uqa}.

\subsection{Background system}
\label{backsubs}

If spatial diffeomorphisms are broken, in the absence of specific symmetries 
we can expect at least a  small degree of background anisotropy during inflation.
This can be induced  by direction-dependent components of the EMT. 
On the other hand, we can impose specific conditions, for example associated 
with symmetries, that reduce or cancel the background anisotropies. In this section, 
we first start discussing the general case in which a small degree of background 
anisotropy is admitted in the metric and in the EMT. We adopt a perturbative scheme 
that allows us to carry on our calculations in a straightforward way, and make
manifest  how to evade Wald's no-hair theorem~\cite{Wald:1983ky} within an 
EFT approach. Then we discuss an example of conditions that allows us to remove  
the background anisotropies, yet maintaining  direction-dependent profiles for the 
fields constituting the background EMT. Hence the system still maintains interesting  
features associated with broken spatial diffeomorphisms, that will become 
important in the next section when discussing quadratic fluctuations and operators
that break discrete symmetries.

During  inflation we consider a background metric that is decomposed as 
homogeneous FRW and anisotropic parts, denoted by $\bar{g}_{\mu\nu}^{(0)}$ and
$\bar{g}_{\mu\nu}^{(a)}$ respectively, as
\begin{equation}
\label{backmet}
\bar{g}_{\mu\nu} = \bar{g}_{\mu\nu}^{(0)} + \bar{g}_{\mu\nu}^{(a)} = a^2(\eta)
\begin{pmatrix}
-1 & \\
& \delta_{ij}
\end{pmatrix}
+ a^2(\eta)
\begin{pmatrix}
& \beta_i \\
\beta_i & \chi_{ij}
\end{pmatrix} \, ,
\end{equation}
where $\beta_i$ and $\chi_{ij}$ are transverse and traceless. We assume from 
now on that $\beta_i$ and $\chi_{ij}$ are small: $|\beta_i|\ll1$ and $| \chi_{ij}|\ll1$.
Hence we consider their contributions only at linearized order in our analysis. 
In other words, we develop a perturbative scheme in terms of the small quantities
parameterizing the background anisotropy in the metric and (as we shall see  
in a moment) in the EMT. We stress that \eqref{backmet} is our background metric. 
On top of it, we will include  inhomogeneous perturbations in the next sections.

We now consider a background EMT that is able to support our small deformation
\eqref{backmet} of a FRW background metric. For this aim, we start introducing  
the following anisotropy parameters that will enter in the EMT:
\begin{itemize}
 \item[-]
  A vector $\theta_i$, selecting a preferred spatial direction. 
  \item[-]
  A shear $\sigma_{ij}$, a symmetric, traceless tensor.
\end{itemize}
To be consistent with the fact that the magnitudes of the anisotropic metric 
components $\beta_i$ and $\chi_{ij}$ are small, we assume both these anisotropic 
parameters $\theta_i$ and $\sigma_{ij}$ to be  small, and treat them at linearized  
order in our discussion. We can think of these objects as {\it vevs} of some fields, 
and in realistic cosmological situations at least a mild coordinate dependence is expected.
Since we are implementing an EFT approach to describe  inflationary fluctuations, 
we do not need to specify an underlying theory that provides such quantities, 
and the equations of motion for the fields associated with them. This since by hypothesis  the 
perturbations of EMT can be set zero by an unitary gauge choice, and do not influence
the dynamics of the metric perturbations on which  we are focusing our attention.  
We only need to ensure that the EMT constructed using these quantities satisfies 
the Einstein equations, order by order in a perturbative expansion in the fluctuations.  
    
In the spirit of the EFT of inflation, 
 the matter action that controls the background
 EMT   breaks both time and space reparametrization invariance, and it is 
then written as
\begin{equation}
S_m = -\int d^4x \sqrt{-g} 
\left[ \Lambda(\eta) + c_1(\eta) g^{00} + c_2(\eta)\delta_{ij}g^{i j}
+ d_1(\eta) \theta_i g^{0i} + d_2(\eta) \sigma_{ij} g^{ij} \right] \, .
\label{linact}
\end{equation}    
 Notice the presence of terms depending on $g^{ij}$ and $g^{0i}$, that are absent in the EFT
 standard where spatial diffeomorphisms are preserved. 
Since the degree of anisotropy is assumed to be small, in what follows we only 
consider contributions at most linear in $\theta_i$ and $\sigma_{ij}$, and in the metric 
deformations $\beta_i$ and $\chi_{ij}$. Moreover, we neglect the possible 
spatial dependence of the coefficients in the previous action. 
The background EMT associated with the action \eqref{linact} is
\begin{equation}
T_{\mu\nu} = -\frac{2}{\sqrt{-g}} \frac{\delta S_m}{\delta g^{\mu\nu}} \, .
\end{equation}
Combined with the Einstein tensor $G_{\mu\nu}$  -- which can be constructed 
straightforwardly from \eqref{backmet} -- the Einstein equations impose the  
following relations to be satisfied at the background level, in a linearized expansion 
for the anisotropy parameters (from now on we set the Planck mass $M_{Pl}=1$):
\begin{align}
3 {\cal H}^2 & = c_1 + 3c_2 + a^2 \Lambda \, ,
\\
\calH^2 - \calH' & = c_1 + c_2 \, , 
\label{eqfc1}
\\
d_1 \theta_i & = c_2\beta_i \, ,
\label{eqefd1}
\\
2 d_2 \sigma_{ij} & =  {\cal H}{\chi}'_{ij} + \frac12 {\chi}''_{ij} + 2c_1 \chi_{ij} \, .
\label{eqfd2}
\end{align}
So we learn that in our linearized approximation the background quantity $\beta_i$ in
the metric is controlled by $d_1$ and the vector $\theta_i$, while $\chi_{ij}$ is 
controlled by $d_2$  and the shear $\sigma_{ij}$. 
A configuration that solves these equations can lead to a solution with a small degree 
of anisotropy in the background during a quasi-de Sitter inflationary stage. 
By choosing appropriately the anisotropic parameters $\theta_i$ and $\sigma_{ij}$
such a configuration can avoid Wald's no-hair theorem~\cite{Wald:1983ky}
and lead to anisotropic inflation, as we discuss in Appendix~\ref{app-wald} using 
the language of the EFT of inflation.

Hence, as a matter of principle, our approach based on the EFT can accommodate
a model-independent analysis of inflationary models with anisotropic backgrounds 
(see e.g.~\cite{anisoBGmodels} for specific models with these properties).

On the other hand, the general analysis of such system can be very cumbersome, due
to several new operators that can contribute.  
For the rest of this article, we make some additional simplifying assumptions
to remove the background anisotropies and facilitate  as much as we can our 
analysis of fluctuations, yet covering some relevant features that are distinctive of 
our system with broken spatial diffeomorphism invariance. 
Our requirements are as follows:
\begin{enumerate}
 \item 
  We impose a residual symmetry~\cite{Dubovsky:2004sg},
  \begin{equation}\label{simimp}
  x^i \to x^i + \xi^i(t)
  \end{equation}
  for an arbitrary  time-dependent function $\xi^i$. Notice that this symmetry 
  invariance is {\it less restrictive} than spatial diffeomorphism, see~\eqref{fulldi}. 
  In our context, this residual symmetry is quite powerful. Since the $0i$ component 
  of the metric perturbation transforms non-trivially under this symmetry 
  (see next section), this symmetry eliminates it from our action, if there are no 
  spatial derivatives acting on it. This requires to choose the parameter $d_1=0$ in 
  the action~\eqref{linact}, and consequently~\eqref{eqefd1} tells us that the metric
  anisotropic parameter $ \beta^i$ vanishes:
  \begin{equation}
  \beta^i = 0 \, .
  \end{equation}
 \item
  In addition, from now on we set the shear equal to zero,
  \begin{equation}
  \sigma_{ij} = 0 \, ,
  \end{equation}
  and focus on the effects of the vector $\theta_i$ only. Setting the shear to zero 
  implies a vanishing source in~\eqref{eqfd2} for the  background anisotropic tensor 
  $\chi_{ij}$. For simplicity, in what follows we   choose the solution
  corresponding to the configuration, 
  \begin{equation}
  \chi_{ij} = 0 \, .
  \end{equation}
\end{enumerate}
After imposing these two requirements we obtain an {\it isotropic} and homogeneous 
FRW background metric. However, the anisotropic parameter $\theta_i$ contributing 
to the background EMT can be non-vanishing, and as we shall see next it can play 
an important role to characterize quadratic operators that break discrete symmetries, 
in the quadratic action for perturbations.

\section{Quadratic operators that break discrete symmetries}
\label{sec-qua}

In this section we discuss how to build a quadratic Lagrangian for the metric 
fluctuations in our setup. We mainly concentrate on operators that break discrete symmetries 
during inflation. We work within the generalized unitary gauge context explained in
Section~\ref{sec-sf}, and consider an homogeneous and isotropic background. 
The operators that we consider in this section are a selection chosen for the most notable
phenomenological consequences. We stress that higher derivative symmetry breaking 
operators -- even preserving spatial diffeomorphisms -- can also be included, but ours are 
the leading ones in a derivative expansion given our symmetry choices.  We make use of the background vector 
$\theta_i$ introduced in the previous section for constructing our quadratic operators, 
and we work at linearized order on this small quantity.

The linearized perturbations around our isotropic background, 
$g_{\mu\nu} = \bar{g}_{\mu\nu} + a^2(\eta)h_{\mu\nu}$,  
can be decomposed into scalar, vector, and tensor sectors:
\begin{align}
h_{00} & = 2A \, ,
\\
h_{i0} & = S_i + \partial_i B \, ,
\\
h_{ij} & = 2\varphi\delta_{ij} + 2\partial_i\partial_jE
+ \partial_iF_j + \partial_jF_i + \gamma_{ij} \, .
\end{align}
Under the most general diffeomorphism transformations \eqref{fulldi}, 
the quantities that appear in the decomposition of $h_{\mu\nu}$ transform as, 
{\em with $\beta^i = \sigma_{ij} = \chi_{ij} = 0$}~\cite{Noh:2004bc},
\begin{align}
A & \to A - \partial_\eta\xi^0 - \calH\xi^0 \, ,
\\
S_i & \to S_i - \partial_\eta\xi_i^T \, ,
\\
B & \to B - \partial_\eta\xi^L + \xi^0 \, ,
\\
\varphi & \to \varphi - \calH\xi^0 \, ,
\\
E & \to E - \xi^L \, ,
\\
F_i & \to F_i - \xi_i^T \, ,
\\
\gamma_{ij} & \to \gamma_{ij} \, .
\end{align}
As we explained, our setup breaks both space and time diffeomorphism 
invariance, but we impose invariance under the residual symmetry transformation 
of \eqref{simimp} that ensures that the quadratic action for the metric 
perturbations does not contain contributions proportional to the metric 
components $h_{0i}$, if there are no spatial derivatives acting on them.

In Appendix~\ref{applist} we list the new derivative operators that are allowed by  
the previous requirements. Here, after discussing the Einstein-Hilbert action and 
the leading operators that do not contain derivatives -- the mass terms -- we 
concentrate on derivative operators that break discrete symmetries. 
Our derivative operators can be considered as leading derivative corrections
to the mass terms that break spatial diffeomorphisms and discrete symmetries in 
Lorentz violating theories of massive gravity~\cite{Dubovsky:2004sg,LVmassivegrav}.

\subsubsection*{$\blacktriangleright$ Einstein-Hilbert action and mass terms}

We start with the Einstein-Hilbert action for quadratic fluctuations. Once  
decomposed into scalar, vector and tensor parts, they read respectively as
follows~\cite{Mukhanov:1990me}:
\begin{align}
S^{(s)} & = \int d^4x \frac{a^2}{2} \left[ -6\left( \varphi' - \calH A \right)^2 
- 2(2A+\varphi)\nabla^2\varphi 
+ 4\left( \varphi'-\calH A \right)\nabla^2\left(B-E'\right) \right] \, ,
\\
S^{(v)} & = \int d^4x a^2 \left[-(S_i-F'_i)\nabla^2(S_i-F'_i) \right] \, ,
\\
S^{(t)} & = \int d^4x \frac{a^2}{8} \left[ {\gamma_{ij}'}^2 - (\nabla\gamma_{ij})^2 \right] \, .
\end{align}
Repeated spatial indices are contracted with $\delta_{ij}$.

To this action we can include the mass operators that are allowed by our 
symmetries:
\begin{align}
{\cal O}_1^{(0)} & = -m_1^2a^4 h_{ij}^2 = 
-m_1^2a^4 \left[ 12\varphi^2 + 2 \left( \partial_i F_j \right)^2 + \gamma_{ij}^2 
+ 8\varphi\nabla^2{E} + 4(\nabla^2{E})^2 \right] \, ,
\label{mass01}
\\
{\cal O}_2^{(0)} & = -m_2^2a^4 h_{ii}^2 =
-m_2^2a^4 \left( 6\varphi + 2\nabla^2{E} \right)^2 \, ,
\\
{\cal O}_3^{(0)} & = -m_3^2a^4 h_{00}^2 = -m_3^2a^4 (4A^2) \, ,
\\
{\cal O}_4^{(0)} & = -m_4^2a^4 h_{00}h_{ii} = 
-m_4^2a^4 \left( 12A\varphi + 4A\nabla^2{E} \right) \, .
\end{align}
These are the zero-derivative [hence the superscript $(0)$], leading operators 
that break diffeomorphism invariance. These operators, and the ones that we  
meet next, already contain the square root of the metric, and can be included  
as they stand into the action. For example the operator \eqref{mass01} can be 
included in the action  as 
\begin{equation}
\Delta S_1^{(0)} = \int d^4x {\cal O}_1^{(0)} \, .
\end{equation}
These mass terms can lead to a non-vanishing anisotropic EMT, 
that among other things does not respect the adiabaticity condition and leads to 
non-conservation of the curvature perturbation on super-horizon scales. 
See~\cite{Cannone:2014uqa} for a discussion on this point.

\subsubsection*{$\blacktriangleright$ Single-derivative operators}

We now consider some novel single-derivative operators, built with or without   
the anisotropic vector $\theta_i$, that have the feature to break discrete 
symmetries in scalar and/or tensor sectors. As discussed in the introduction, 
there is a rich literature on possible interactions that violate the discrete parity 
symmetry, and their consequences for the CMB. The novelty of our 
model-independent approach is the use of EFT for inflation in a context where  
spatial diffeomorphism invariance can be explicitly broken (see 
also~\cite{Weinberg:2008hq} for a discussion of parity violating operators in 
an EFT for inflation preserving spatial diffeomorphism invariance). As we discussed, 
spatial diffeomorphism invariance can be violated in inflationary systems where 
background fields acquire spatial-dependent background values, as in models with 
vectors or in solid inflation. If discrete symmetries are not imposed {\it a priori}, 
the operators that we consider can be expected to be generated by quantum effects 
in such inflationary scenarios. For this reason, it is interesting to explore them and 
their consequences.   
Here we introduce a couple of such operators, the ones with the most notable
phenomenological consequences that will be studied in the next section.

The lowest dimensional, single derivative  operator that breaks parity does not involve 
anisotropic parameters and reads
\begin{equation} \label{paruno}
{\cal O}_1^{(1)} \,=\,\mu\,  a^3
\epsilon_{ijk} \left( \partial_i h_{jm} \right) h_{km}
\,=\,\mu\, a^3
\epsilon_{ijk} \left[
\left( \partial_i \gamma_{jm} \right) \gamma_{km}
- \partial_i F_{j} \nabla^2 F_k
\right] \, .
\end{equation}
It leads to {\em parity violation} in the tensor sector, since it is not invariant   
under the interchange $x^i \to -x^i$. $\mu$ is a mass scale we have included for dimensional reasons.

In addition, there is another interesting   
single-derivative operator, built with the background vector $\theta_i$, 
that contains a single derivative along time:
\begin{align}
{\cal O}_2^{(1)} & = \, \mu\,
a^3 \epsilon_{ijk} \theta_i h_{jm} {h}'_{km}
\nonumber\\
& =\, \mu\, a^3 \epsilon_{ijk} \theta_i \left( \gamma_{jm} {\gamma}'_{km}
- F_m \partial_j \gamma'_{km} - F'_m \partial_k \gamma_{jm} - F_j \nabla^2 F'_k
+ 2\partial_j F'_k \nabla^2 E + 2 \partial_k F_j \nabla^2 E' \right) \, . 
\label{pardue}
\end{align}
We can say that such an operator breaks {\em time-reversal} in the tensor sector,  
since the contributions within the parenthesis are not invariant under a change of 
sign in the time direction. Notice that in order to build it we need to use the vector
$\theta_i$ that selects a preferred direction. Recent papers discussed possible 
phenomenology of scenarios that contain together background anisotropies and 
parity violation: see e.g.~\cite{BGaniso-PV}. We will see that such an operator can 
have interesting consequences for the dynamics of the tensor modes.

\subsubsection*{$\blacktriangleright$ Two-derivative operators}

Among the possible two-derivative operators, we focus on  two interesting   
ones that break discrete symmetries:
\begin{align}
{\cal O}_1^{(2)} & = a^2 h'_{ij} \theta_j \partial_k h_{ik}
\nonumber\\
& = -a^2 \theta_j \Big(
4\varphi' \partial_j \varphi + 2\varphi'\nabla^2F_j + \gamma_{ij}'\nabla^2 F_i
 - 2\varphi\nabla^2F_j' - 2 F'_j \nabla^4 E
\nonumber\\
& \hskip4.5em
+ 4\varphi'\partial_j\nabla^2E + 4\partial_j\varphi\nabla^2 E'
+4 \nabla^2 E'  \partial_j\nabla^2 E
- F_i'  \partial_j \nabla^2 F_i \Big) \, ,
\label{nopsc}
\\
\label{nopsc2}
{\cal O}_2^{(2)} &= a^2 h'_{ij} \theta_k \partial_k h_{ij}
\nonumber\\
& = -a^2 \theta_k \Big( 
12\varphi'\partial_k\varphi + \gamma_{ij}'\partial_k\gamma_{ij}
\nonumber\\
& \hskip4.5em
+ 4\varphi'\partial_k\nabla^2E + 4\partial_k\varphi\nabla^2 E'
+ 4\nabla^2E'\partial_k\nabla^2E - F'_{i} \partial_k \nabla^2 F_{i} \Big) \, .
\end{align}
Notice that, considering their scalar and tensor parts, such operators are 
not invariant under an (independent) interchange of spatial and of time 
coordinates. Hence we can say that these operators break both parity and 
time-reversal, in the tensor as well as in the scalar sectors. In the next section
we will discuss their consequences.

Other single and two derivative operators that can break discrete symmetries 
are listed in Appendix~\ref{applist}.

\section{Dynamics of linearized fluctuations}
\label{sec-lin}

We now discuss some consequences of the discrete symmetry breaking operators
that we presented in the previous section. We concentrate our attention to the 
dynamics of linearized fluctuations. Within our approximation of small anisotropy 
parameter $\theta_i$, we show that vector degrees of freedom do not propagate. 
Scalar degrees of freedom acquire a direction-dependent phase. Although this phase 
factor does not have consequences for the scalar power spectrum, nevertheless 
it might affect higher order correlators. We also show that small direction dependent
contributions to the sound speed can arise. 
 At the quadratic level, the most notable 
consequences occur in the tensor sector, where we find that some of our new 
operators lead to a chiral amplification of gravity waves. This is more effective than 
the one first pointed out in~\cite{Lue:1998mq} discussing parity-breaking operators, 
because the modes can be  continuously amplified during the whole inflationary epoch.

\subsection{No propagating vector modes}

At linear order in the anisotropy parameter $\theta_i$, we can arrange our system  
such  that there are no propagating vector degrees of freedom: the derivative operators
of the previous section have been selected, among other things,   to ensure this condition. To see this, 
we include for simplicity a single mass term, proportional to $m_1^2$, as given by 
\eqref{mass01}, plus a combination of the discrete symmetry breaking operators 
proportional to $\theta_i$ that we have introduced in the previous section. 
The quadratic vector Lagrangian can be expressed as
\begin{equation}\label{lagV}
{\cal L}^{(v)} = \frac{a^2}{2} \left[ \partial_k \left( S_i - F_i'\right) 
\partial_k \left( S_i - F_i' \right) \right] - 2m_1^2a^4 \left( \partial_i F_j \right)^2
- \theta_i F_j \left( \cdots \right) \, ,
\end{equation}
where the dots contain contributions depending on $F_k$ or on scalar and 
tensor fields, that we do not need to specify for our arguments. In the previous 
expression, the first part comes from the Einstein-Hilbert term, the second from 
a mass term, while the third part collects the contribution from the new derivative 
operators discussed in the previous section. In this context,  the vector $S_i$ 
appears only in the first term of \eqref{lagV}. It  can be readily integrated out, 
leaving a Lagrangian identical to \eqref{lagV} but with the first term missing. 
The equation of motion for $F_i$ then reads
\begin{equation}\label{condV}
\nabla^2 F_i = \frac{\theta_i}{m_1^2} \left( \cdots \right) \, ,
\end{equation}
where again the dots contain contributions of the various fields involved, that 
we do not need to specify. Substituting \eqref{condV} into \eqref{lagV}, we find 
only terms of $\calO(\theta_i^2)$ that are negligible within our approximation. 
Hence, although typically vector modes propagate in our context, at linearized 
order in $\theta_i$, the vector degrees of freedom are not dynamical and will 
be set to zero.

\subsection{Direction-dependent phase in the scalar sector}

Let us examine the effects of parity breaking and time-reversal operators in the 
scalar sector. We  consider a quadratic action built in terms of the Einstein-Hilbert
contributions,  mass terms, and a linear combination of the two-derivative operators
${\cal O}_i^{(2)}$ introduced in \eqref{nopsc} and \eqref{nopsc2}. We set the vector
perturbations to zero as seen in the previous subsection.  This scalar action contains 
four scalar degrees of freedom: $A$, $B$, $E$ and $\varphi$. Among them, $A$ and 
$B$ are non-dynamical and can be integrated out, leaving a scalar Lagrangian for 
$E$ and $\varphi$. We can proceed as done in \cite{Cannone:2014uqa}, further solve 
the equation of motion for the non-dynamical field $E$, and plug it into the action. 
We find at linearized order in $\theta_i$ a   two-derivative operator for 
$\varphi$ (already present in our expressions for ${\cal O}_i^{(2)}$)
 that breaks the discrete parity and time-reversal symmetries:
\begin{equation} 
\label{newso1}
\calL^{(s)} \supset a^2 \varphi' \theta_i \partial_i \varphi \, .
\end{equation}
Other contributions quadratic or higher in the parameter $\theta_i$ can
 be neglected, as done in the previous subsection for vector fluctuations. 
The scalar field $\varphi$ in the unitary gauge is the curvature perturbation $\calR$,
hence its statistics can be directly connected with observable quantities. 
Here however we limit our attention to understand how the operator \eqref{newso1} 
modifies the mode function for $\varphi$, viz. $\calR$. We consider then the action 
for the canonically normalized field $u = z\calR$ with $z \propto a$ during 
quasi-de Sitter expansion:
\begin{equation}
\label{eq:scalaraction}
S^{(s)} = \int d^4x \frac12 \left[ {u}'^2 - \left(\nabla { u}\right)^2 
+ \frac{z''}{z} u^2 + 2{b}_1\theta_i{u}' \partial_i{u} \right] \, ,
\end{equation}
where the last is our new term, weighted by a real  coefficient $b_1$ that for simplicity 
we consider as constant. Here we do not explicitly discuss the consequence of the
mass terms ${\cal O}^{(0)}_{i}$. Such contributions have been already studied 
for example in~\cite{Cannone:2014uqa} and have been shown to lead to anisotropic 
stress and non-conservation of the curvature perturbation, generalizing the results 
first pointed out for solid inflation~\cite{Endlich:2012pz}.

The equation of motion for the mode function $u_k$,  that follows from 
\eqref{eq:scalaraction} once converted to Fourier space, results 
\begin{equation}
\label{equk1}
u_\mathbi{k}'' + 2ib_1\theta_ik_iu_\mathbi{k}' + \left( k^2 - \frac{z''}{z} \right) u_\mathbi{k} = 0 \, .
\end{equation}
 At early times the new operator proportional to $b_1$ is subdominant,  
so a standard Bunch-Davies vacuum can be unambiguously defined. 
 It is convenient to express the mode function $u_\mathbi{k}$ as
\begin{equation}\label{solph1}
u_\mathbi{k} =
  e^{-i\theta_ik_ib_1\eta} 
u_k^{(0)}
\end{equation}
so that \eqref{equk1} becomes
\begin{equation} \label{eqzmo}
{u^{(0)}_{k}}'' + \left( k^2 - \frac{z''}{z} \right) u^{(0)}_{ k} 
+ k^2 \left( b_1\theta_i\hat{k}_i \right)^2 u^{(0)}_{ k}  = 0 \, ,
\end{equation}
where $\hat{k}_i \equiv k_i/k$. 
The last term in the previous expression is quadratic in $\theta_i$, so it
can be neglected for consistency with our approximation (but see the comment
at the end of this subsection). Doing so we end with the standard evolution 
equation in a FRW background, and the solution 
for $u^{(0)}_{k}$ can be expressed in terms of  Hankel functions. On top of this, 
the complete solution for $u_\mathbi{k}$ gains a new direction-dependent 
contribution to the phase proportional to $b_1$ as in \eqref{solph1}.
Such a configuration is only reliable at linearized order in $\theta_i$, hence on 
large scales, $k/(aH)\le1$. For smaller scales, contributions that are non-linear
in $\theta_i$ can become large and change the solution:  this fact is important 
when quantizing the system. Note that the power spectrum remains isotropic 
because \eqref{solph1} is different from the standard solution by a 
direction-dependent phase, which cancels when computing the power spectrum.

It is also interesting to interpret the role of this phase in coordinate space, making 
a Fourier transform of \eqref{solph1}. One finds that 
\begin{equation}
u(\eta,x^i) = u^{(0)}(\eta,x^i + b_1\,\eta\,\theta^i) \, .
\end{equation}
Hence its effect amounts to a time-dependent shift of the argument of the scalar 
mode function in coordinate space. Such shifts cancel when taking correlation 
functions among scalar fluctuations, due to the translational invariance of these 
quantities. On the other hand, they can have non-vanishing physical effects when 
taking higher order correlations functions between scalar and tensor modes, since 
the tensor perturbations do not necessarily share the same shifts. It would be 
interesting to study this topic further.

Let us end this subsection briefly commenting on the last term in 
 \eqref{eqzmo}: as we explained above, consistency of our approximations
 would require to neglect such terms, since at quadratic order in the anisotropy
 parameter $\theta_i$ other contributions of comparable
 size can arise -- for example 
 the coupled terms between scalar, vector and tensor fluctuations 
 -- that should
 be taken into account.  
 Nevertheless, such a particular term would be present, and provide a quadrupole contribution
 to the scalar sound speed. It would be interesting to study its  effects, noticing also that being
 of positive size it {\it increases} the amplitude of the sound speed rendering it larger
 than one. 
We leave the analysis of this topic  to future work.

\subsection{Chiral phase in the tensor sector}

We now explore the consequences of our discrete symmetry breaking operators 
for the  tensor sector. We do not consider the mass terms, which were studied e.g.
in~\cite{Cannone:2014uqa}. We consider here the effects of the single derivative 
operators ${\cal O}^{(1)}_{1}$ and $\calO^{(1)}_2$ that   break parity and time-reversal. 
The consequences of the two-derivative operator ${\cal O}^{(2)}_{i}$ are, as can be read 
from the derivative structure of the tensor perturbations in \eqref{nopsc2}, identical to the 
ones discussed in the previous section on the scalar sector, so we do not analyze them here.

The action for tensor fluctuations is
\begin{equation}
S^{(t)} = \int d^4x \frac{a^2}{8} \left[ {\gamma_{ij}'}^2 - \left( \nabla\gamma_{ij}\right)^2 
+ 2q_1\,\mu\,a\,\epsilon_{ijk} \left( \partial_i\gamma_{jm} \right)\gamma_{km} 
+ 2q_2\,\mu\,a\,\epsilon_{ijk}\theta_i\gamma_{jm}\gamma_{km}' \right] \, ,
\end{equation}
where we have assumed the the coefficients $q_1$ and $q_2$ are constant dimensionless real 
parameters: the condition of being real is  required by our conventions on the tensor
polarizations. The equation of motion for the tensor degrees of freedom results
\begin{equation}\label{teneqbc}
\gamma_{ij}'' + 2{\cal H}\gamma'_{ij} - \nabla^2\gamma_{ij} 
- 2{q_1\,\mu\,a\,} \epsilon_{kmi} \partial_k \gamma_{mj}
+ 2{q_2\,\mu\,a\,} \epsilon_{kmi} \theta_k \gamma'_{mj}
+ 3{q_2}\,\mu\,a\,{\cal H}\, \epsilon_{kmi} \theta_k \gamma_{mj} = 0 \, ,
\end{equation}
where all indices are contracted with the Kronecker symbol $\delta_{ij}$. Now, we 
introduce the circular polarization tensor ${\bf e}_{ij}^{(\lambda)}(\hat{\mathbi{k}})$, 
with $\lambda = +$ $(-)$ corresponding to the right (left) circular polarization, 
which satisfies the circular polarization conditions (see Appendix~\ref{App-Circular}):
\begin{equation}\label{circpol}
\begin{split}
{\bf e}^{(\lambda)}_{ij}k_j & = {\bf e}_{ii}^{(\lambda)} = 0 \, ,
\\
\epsilon_{ilm}{\bf e}_{lj}^{(\lambda)}k_m & = i\lambda k {\bf e}_{ij}^{(\lambda)} \, ,
\\
{\bf e}_{ij}^{(\lambda)*} {\bf e}_{ij}^{(\lambda')} & = 2 \delta_{\lambda\lambda'} \, .
\end{split}
\end{equation}
$\gamma_{ij}$ can be Fourier expanded in terms of polarization mode functions as
\begin{equation}
\gamma_{ij}(\eta,\mathbi{x})
= \int \frac{d^3 k}{(2 \pi)^3}\,\sum_{\lambda=+,\,-}\,
\left[\gamma_{(\lambda)}(\eta,\mathbi{k})
{\bf e}_{ij}^{(\lambda)}(\hat{\mathbi{k}})
e^{i\mathbi{k}\cdot\mathbi{x}}
+h.c.\right] \, .
\end{equation}
Then, we find the equation of motion for the mode function $\gamma_{(\lambda)}$, 
after contracting with ${\bf e}_{ij}^{(-\lambda)}$, as
\begin{equation}
\gamma_{(\lambda)}'' + 2 {\cal H}\gamma_{(\lambda)}' + {k^2}\gamma_{(\lambda)}
- 2\lambda q_1 \,\mu\,a\,{k}\gamma_{(\lambda)}
- 2\lambda q_2 \,\mu\,a\, \hat{\theta} \gamma_{(\lambda)}'
- 3\lambda{q_2\,\mu\,a\,{\cal H}}{\hat{\theta}}\gamma_{(\lambda)} = 0 \, ,
\end{equation}
where we have introduced 
\begin{equation} 
\label{defth}
\hat{\theta} \equiv \frac{\lambda}{2} {\bf e}^{(\lambda)}_{ij} \epsilon_{lmi}
\theta_l {\bf e}_{mj}^{(-\lambda)} \, .
\end{equation}
Let us discuss the  interpretation\footnote{We thank Angelo Ricciardone and Maresuke Shiraishi for pointing out this argument to us.
} of $\hat{\theta}$. Using \eqref{deftpol}, we learn that
$\hat{\theta} \,=\, {\lambda} { e}^{(\lambda)}_{i} \epsilon_{lmi}
\theta_l { e}_{m}^{(-\lambda)}$. Here  $ { e}^{(\lambda)}_{i} $ and $ { e}_{m}^{(-\lambda)}$
are two mutually  orthogonal vectors, that are both orthogonal to the direction 
of the three-momentum $\mathbi{k}$. This implies that the cross product 
${ e}^{(\lambda)}_{i} { e}_{m}^{(-\lambda)}\, \epsilon_{mil}$ 
is a vector parallel to $\mathbi{k}$: contracting it
with the vector $\theta_l $ and using \eqref{circpol} leads to the identity
\begin{equation} 
\label{defth2}
\hat{\theta} = i{\theta}_i\hat{k}_i \,. 
\end{equation}
Notice at this stage the main difference between the operators proportional
to $q_1$ and $q_2$. The operator proportional to $q_2$
is associated with time-derivatives of the mode function $\gamma_{(\lambda)}$ 
or the scale factor, while the operator $q_1$ with space-derivatives. 
The effect of the contribution of $q_1$ corresponds to the known parity-violating 
operators~\cite{Lue:1998mq}, and produces an enhancement/suppression of 
tensor mode polarization at horizon crossing only. Such effects are well studied 
in the literature (see as an example the review~\cite{Maleknejad:2012fw}) so 
we will not study them here. Let us instead concentrate on the consequences 
of the novel operator ${\cal O}_2^{(1)}$ proportional to $q_2$.
We rescale the field $\gamma_{(\lambda)}$ in the standard manner as
\begin{equation}
v_{(\lambda)} \equiv \frac{a}{\sqrt{2}}\gamma_{(\lambda)} \, .
\end{equation}
The equation of motion for $v_{(\lambda)}$ is then
\begin{equation}
v_{(\lambda)}'' - 
2i\lambda q_2 \mu a \theta_i\hat{k}_i v_{(\lambda)}'
+ \left( k^2 - \frac{a''}{a} - 
i\lambda q_2 \mu a\calH \theta_i\hat{k}_i  
\right) 
v_{(\lambda)} = 0 \, .
\end{equation}
Similar to what we did for the scalar sector, it is convenient to rescale 
\begin{equation}
v_{(\lambda)} \equiv
e^{
 i\lambda q_2\mu\theta_i\hat{k}_i \int a d\eta
}
v_{(\lambda)}^{(0)} \, .
\end{equation}
The equation for $v_{(\lambda)}^{(0)}$, at linear order in $\theta_i$ and
so neglecting quadrupolar effects,
reduces to the well-known form
\begin{equation}
{v_{(\lambda)}^{(0)}}'' + \left( k^2 - \frac{a''}{a} \right) v_{(\lambda)}^{(0)} = 0 \, .
\end{equation}
This equation is identical to the standard mode function equation for the tensor 
perturbations. Furthermore, we notice that, neglecting slow-roll corrections, we can write 
\begin{equation}
\int a d\eta = \frac{N_e}{H} \, ,
\end{equation}
with $N_{e}$ being the number of $e$-folds, and $H$ the value of the Hubble parameter 
during inflation. 
So the solution for $\gamma_{(\lambda)}$ is given by
\begin{equation}
\gamma_{(\lambda)} \,=\,
\exp \left( i\lambda q_2 \mu \theta_i \hat{k}_i \frac{ N_{e}}{H} \right) 
\, \gamma_{(\lambda)}^{(0)} \, .
\end{equation}
Therefore, we
again find a phase modulation of the wavefunction -- like in the scalar sector -- but
now the coefficient of this phase depends on the chirality of the specific gravity wave one is 
considering, and  on the number of $e$-folds \textcolor{black}{as well}. Such \textcolor{black}{a} phase does not influence the power
spectrum, since 
 it can be read as a ``chiral'' translation of the modes when expressed in \textcolor{black}{the}  coordinate
space:
\begin{equation}
\gamma_{(\lambda)}(\eta,x^i)
= \gamma_{(\lambda)}^{(0)}\left(\eta,x^i-
\lambda q_2\mu \frac{ N_{e}}{H} \theta^i
\right) \, .
\end{equation}
A translation in the coordinates does not affect the power spectrum of the tensor modes, since
the power spectrum   is translationally  invariant. On the other hand, since the translation depends
on the chirality, it can affect the bispectra among tensor modes with different chirality (as studied
for example in \cite{grnG}), as well as bispectra between tensor and scalar sectors. We hope to 
return to investigate these topics in the near future.

\section{Conclusions}\label{sec-con}

The EFT of inflation has emerged in recent years as a powerful unifying approach 
for analyzing broad classes of inflationary scenarios. A systematic analysis of 
effective operators compatible with the  symmetries assigned to a system might  
suggest new effects, not predicted or analyzed within the specific models studied 
so far, that if supported by observations can motivate new directions for model
building.

In this article we have studied the consequences of breaking discrete symmetries 
during inflation using the model-independent language of EFT. We have developed 
aspects of EFT of inflation in a context where all diffeomorphisms are broken 
during the inflationary phase. We have shown that this effective description allows 
one to describe systems where background spatial anisotropies can be present during 
inflation. Moreover, we have discussed how to avoid Wald's no-hair theorem within 
the model-independent language of EFT of inflation. Then we have  focussed on studying  
the leading operators at the quadratic order in perturbations that  break discrete 
symmetries. We have identified operators that break parity and time-reversal
during inflation, 
and analyzed their consequences for the dynamics of linearized fluctuations. 
Both in the scalar  and tensor sectors, we show that such operators can lead to a new 
direction-dependent phase for  modes involved. Such a directional phase does not 
affect the power spectrum, but can  have consequences for higher correlation 
functions. Moreover, a small quadrupole contribution to the  sound speed
can be generated.
We stress that  using an EFT approach  we did not have to commit on 
specific models to develop our arguments, that are based on general features 
of the system of fields driving inflation.

Our investigations can be further elaborated in various directions. At theoretical 
level, it would be interesting to extend our model-independent approach to derive 
a third order action for perturbations, and study how new discrete symmetry 
breaking operators and direction-dependent effects can influence non-Gaussian 
observables. After analyzing novel ramifications of such findings, it will also be 
important to study actual realizations and concrete models able to generate 
the new operators that break discrete symmetries in the manner  studied here.
At observational level, it would be interesting to study distinctive consequences of 
our discrete symmetry breaking operators for the properties of the CMB, in particular 
for what respect specific correlations between $T$, $E$ and $B$ modes, at the level 
of two- and three-point functions.  We leave these interesting questions to future study.

\subsection*{Acknowledgments}

It is a pleasure to thank Marco Bruni, Hassan Firouzjahi, Eichiiro Komatsu, Azadeh Maleknejad, David Wands, Masahide Yamaguchi and especially Angelo Ricciardone and Maresuke Shiraishi  for helpful discussions and comments on the draft, 
and the Munich Institute for Astro- and Particle Physics (MIAPP) of the DFG cluster of excellence ``Origin and Structure of the Universe'' for hospitality.
GT is supported by an STFC Advanced Fellowship ST/H005498/1. 
JG acknowledges the Max-Planck-Gesellschaft, the Korea Ministry of Education, Science and Technology, Gyeongsangbuk-Do and Pohang City for the support of the Independent Junior Research Group at the Asia Pacific Center for Theoretical Physics, and is also supported by a Starting Grant through the Basic Science Research Program of the National Research Foundation of Korea (2013R1A1A1006701). 

\newpage

\begin{appendix}

\section{Wald's no-hair theorem in the  EFT of inflation}
\label{app-wald}

In this appendix we discuss how Wald's isotropization theorem~\cite{Wald:1983ky} 
can be violated using  the language of EFT of inflation, justifying the approach 
developed in Section~\ref{backsubs}.

Wald's theorem states that, under some hypothesis on the EMT that we will review 
below, the inflationary expansion rapidly reduces the amplitude of background 
anisotropies to an unobservable level. Here we show that the prerequisites behind 
the theorem are not necessarily satisfied in our case. We use results and notation of 
Section~\ref{backsubs}. We can write
\begin{equation}
\calH' = \calH^2(1-\epsilon) \, , 
\end{equation}
where $\epsilon \equiv -a^{-1}H'/H^2$ with $\calH = aH$. Substituting this result into 
\eqref{eqfc1}, we find that $c_1+c_2\,=\,\epsilon\,{\cal H}^2$. Using this information,  
the background EMT can be decomposed  as 
\begin{equation}
T_{\mu\nu}\,=\,-\Lambda(\eta)\,\bar{g}_{\mu\nu}+T^{(2)}_{\mu\nu}
\end{equation}
with 
\begin{equation}
\begin{split}
T^{(2)}_{00} & = \epsilon\calH^2 + 2c_2 \, ,
\\
T^{(2)}_{0i} & = \left( \epsilon\calH^2 - 2c_2 \right) \beta_i \, ,
\\
T^{(2)}_{ij} & = \left( \epsilon\calH^2 - 2c_2 \right) \left( \delta_{ij} + 3\chi_{ij} \right) 
+ \calH\chi_{ij}' + \frac{1}{2}\chi_{ij}'' \, .
\end{split}
\end{equation}
Wald's isotropization theorem states that anisotropies are rapidly suppressed during 
inflation if the strong and dominant energy conditions are satisfied:
\begin{align}
\left( T^{(2)}_{\mu\nu} - \frac{1}{2}\bar{g}_{\mu\nu}T^{(2)} \right) t^\mu t^\nu & \ge 0 \quad 
\text{for all time-like vectors $t^\mu$} \, ,
\\
\label{doenco}
T^{(2)}_{\mu\nu} \hat{t}^\mu \hat{t}^\nu & \ge 0 \quad 
\text{for all future-directed, causal vectors $\hat{t}^\mu$} \, .
\end{align}
Time-like vectors $t^\mu$ satisfy the condition
\begin{equation}
\left( t^0 \right)^2 \ge 2\beta_it^0t^i + \left( \delta_{ij} + \chi_{ij} \right) t^it^j \, .
\end{equation}
In our case, the dominant energy condition reads
\begin{align}
& \left( \epsilon\calH^2 - 2c_2 \right) \left[ \left(t^0\right)^2 + 2\beta_it^0t^i + 
\left( \delta_{ij}+\chi_{ij} \right)t^it^j \right]
\nonumber\\
& + 4c_2\left(t^0\right)^2 + 2 \left( d_2\sigma_{ij} - c_2\chi_{ij} \right)t^it^j \ge 0 \, .
\end{align}
In EFT scenarios with no breaking of spatial diffeomorphisms or isotropy, $c_2=0$,  
$\sigma_{ij}=0$ and $\chi_{ij}=0$. The second line in the above equation would vanish, 
while the first line would be positive definite, satisfying in this way the dominant energy 
condition \eqref{doenco}. In our more general setup, instead, the second line is 
non-vanishing, and can render the previous quantity negative. Hence, in general the
prerequisites underlying Wald's theorem can be expected to be violated in our context 
based on the EFT of inflation. Such situations can be realized in models of inflation 
with vector fields~\cite{Watanabe:2009ct}, or solid inflation, as discussed in the recent 
literature, see e.g.~\cite{Wald-breaking}.

\section{List of allowed operators}
\label{applist}

In this appendix we list new derivative operators that satisfy our requirements, 
besides the ones already presented in the main text and in~\cite{Cannone:2014uqa}.
To avoid possible ghost pathologies, we do not consider operators that contain time 
derivatives on $h_{00}$ and $h_{0i}$. Moreover, to satisfy the residual symmetry 
\eqref{simimp} we consider operators containing $h_{0i}$ only when spatial
derivatives act on it.

\subsubsection*{\it Single-derivative operators}

The new single-derivative operators are the following:
\begin{equation}
\begin{split}
& h_{0i,i}h_{00} \, , \quad h_{0i,i}h_{jj} \, , \quad h_{0i,j}h_{ij} \, , \quad h_{ii}'h_{00} \, , 
\quad h_{ii}'h_{jj} \, , \quad h_{ij}'h_{ij} \, , \quad \epsilon_{ijk}h_{00,i}h_{jk} \, ,
\\
& \theta_ih_{ij,j}h_{00} \, , \quad \theta_ih_{jj,i}h_{00} \, , \quad \theta_ih_{ij,j}h_{kk} \, , 
\quad \theta_ih_{ij,k}h_{jk} \, , 
\\
& \theta_i\epsilon_{jkl}h_{0j,k}h_{il} \, , \quad \theta_i\epsilon_{ijk}h_{0j,k}h_{ll} \, , 
\quad \theta_i\epsilon_{ijk}h_{0j,l}h_{kl} \, , \quad \theta_i\epsilon_{ijk}h_{0l,j}h_{kl} \, .
\end{split}
\end{equation}
Note that $\theta_ih_{jj,i}h_{kk}$ and $\theta_ih_{jk,i}h_{jk}$ are allowed but can be 
made as total derivatives, thus we have omitted these operators.

\subsubsection*{\it Two-derivative operators}

The new two-derivative operators are:
\begin{equation}
\begin{split}
& h_{0i,i}h_{jj}' \, , \quad h_{0i,j}h_{ij}' \, , \quad \epsilon_{jkl}h_{ij,k}h_{il}' \, ,
\\
& \theta_ih_{00,i}h_{jj}' \, , \quad \theta_ih_{00,j}h_{ij}' \, , \quad \theta_ih_{ij,j}h_{kk}' \, , 
\quad \theta_ih_{ij,k}h_{jk}' \, , \quad \theta_ih_{jj,i}h_{kk}' \, , \quad \theta_ih_{jj,k}h_{ik}' \, , 
\\
& \theta_i\epsilon_{jkl}h_{0j,k}h_{il}' \, , \quad \theta_i\epsilon_{ijk}h_{0j,k}h_{ll}' \, , 
\quad \theta_i\epsilon_{ijk}h_{0j,l}h_{kl}' \, , \quad \theta_i\epsilon_{ijk}h_{0l,j}h_{jl}' \, .
\end{split}
\end{equation}

\section{Circular polarization conditions}
\label{App-Circular}

We explain in more details our conventions for the circular polarization condition 
for tensor perturbations, \eqref{circpol}~\cite{Maleknejad:2012fw,Shiraishi:2012bh}.
The polarization vector $e_i^{(\lambda)}(\hat{\mathbi{k}})$ perpendicular to 
$\hat{\mathbi{k}}$ can be written as
\begin{equation}
e_i^{(\lambda)}(\hat{\mathbi{k}}) = \frac{\hat\theta_i(\hat{\mathbi{k}}) 
+ i\lambda\hat\phi_i(\hat{\mathbi{k}})}{\sqrt{2}} \, ,
\end{equation}
with $\lambda = \pm$. This vector satisfies
\begin{equation}\label{vecpol}
\begin{split}
k_ie_i^{(\lambda)} & = 0 \, ,
\\
e_i^{(\lambda)*}(\hat{\mathbi{k}}) & = e_i^{(-\lambda)}(\hat{\mathbi{k}}) 
= e_i^{(\lambda)}(-\hat{\mathbi{k}}) \, ,
\\
e_i^{(\lambda)*}e_i^{(\lambda')} & = \delta_{\lambda\lambda'} \, ,
\\
\epsilon_{ijl}k_ie_j^{(\lambda)} & = -i\lambda k e_l^{(\lambda)} \, .
\end{split}
\end{equation}
By means of such a polarization vector we can construct the polarization tensor as
\begin{equation}
{\bf e}_{ij}^{(\lambda)} = \sqrt{2}e_i^{(\lambda)}e_j^{(\lambda)} \, . \label{deftpol}
\end{equation}
It is straightforward to prove \eqref{circpol} using \eqref{vecpol}.

\end{appendix}

\end{document}